# Acknowledgment

This is not the final version of the paper.

The format of the paper awaits final modification, the formula used in the paper may have display errors

The illustrations and tables in the paper are only used for space purposes, and are not quoted pictures of real research

This study did not receive funding from any individual or organization

# Volatility-based strategy on Chinese equity index ETF options

Peng Yifeng

CUHK Shenzhen Scholl of Management and Econommics

Abstract

This study examines the performance of a volatility-based strategy using Chinese equity index ETF options. Initially successful, the strategy's effectiveness waned post-2018. By integrating GARCH models for volatility forecasting, the strategy's positions and exposures are dynamically adjusted. The results indicate that such an approach can enhance returns in volatile markets, suggesting potential for refined trading strategies in China's evolving derivatives landscape. The research underscores the importance of adaptive strategies in capturing market opportunities amidst changing trading dynamics.

# Content

Please delete this content for final version, this contents only for locating content when writing

# Background Introduction:

In recent years, China's derivatives markets have experienced rapid advancements with derivatives trading volumes climbing to considerable levels (Hu & Wang, 2022). This influx is attributed to the expansive implementation of regulatory reforms and the market's adaptation to international practices, catering to growing investor demand for diverse financial instruments (Bryan, Yang, & Wang, 2008). Our research capitalizes on these developments and encompasses an exhaustive dataset of daily ETF options traded on the Shanghai Stock Exchange. By commencing with a short-volatility strategy, we align our work with the collective understanding of derivatives usage in market hedging and speculation tools that characterize modern financial markets (Fan & Liu, 2022).

Our short-volatility strategy, grounded in historical antecedents, has, until 2018, demonstrated robust performance by delivering excess returns over conventional buy-and-hold benchmarks—a phenomenon that resonates with the broad acknowledgment of such strategies' prowess in earning premiums during stable market conditions (Fleming, Kirby, & Ostdiek, 2003). However, following 2018, significant market perturbations have led to the annulment of these excess returns as the risk profile of derivative instruments adapted to the new market realities (Cao, Chen, & Li, 2021). Such seismic shifts necessitate an incisive analysis of trading strategies' adaptability and highlight a transformative phase in the operational dynamics of China's derivatives markets (Chen, Lee, & Shih, 2020).

Since 2018, the efficacy of certain investment strategies has begun to show signs of decline, with a notable absence of clear risk-adjusted returns. This phenomenon can be attributed to various market dynamics, including shifts in market volatility (request quote). To address these challenges, this study proposes enhancements to the existing model by employing volatility forecasting techniques, such as volatility momentum and the Generalized Autoregressive Conditional Heteroskedasticity (GARCH) model, which have been proven to be robust in predicting market volatility (Moran & Liu, 2020). By adjusting positions and market exposure based on these forecasts, the model demonstrates significant improvement, particularly in capturing larger gains during market upswings and mitigating losses during downturns (Yue, Zhang, & Tan, 2020). The revised models underscore the potential for volatility-based trading strategies in the context of Chinese equity index options. This approach leverages the predictive power of volatility to inform strategic trading decisions (Sui, Lung, & Yang, 2021). For instance, Yue et al. (2023) highlighted the relationship between market leverage and volatility feedback effects, which are integral to the design of an effective volatility-based strategy. Additionally, Arkorful, Chen, Liu, and Zhang (2020) investigated the implications of speculative and hedging activities associated with option trading on market volatility.

The research presented here aligns with emerging insights from the literature, suggesting that the introduction of options trading can have a significant impact on the volatility of underlying equities, as evidenced in the Chinese stock markets (Guo, Wang, & Fan, 2022). Moreover, the volatility risk premium and the role of index options in shaping trading demands have been extensively studied, offering critical insights into the mechanics of volatility-based trading (Wang & Zhou, 2021; Pan, Liu, & Roth, 2003). In conclusion, this study contributes to the growing body of knowledge on the efficacy of volatility-based trading strategies. With further refinement and practical implementation considerations, these strategies hold promise for real-world applications in the trading of Chinese equity index options (Xiao, Wen, Zhao, & Wang, 2021).

# 1. Introduction about the market - Stock index ETF options market in China [1]

China's option market has developed quickly in recent years. Though currently standardized options on individual stocks are still not available in China, there are several options based on financial underlying that can be traded through the central exchange (Shanghai and Shenzhen Stock Exchanges), namely options on equity index ETF such as 50ETF (510050) and 300ETF (510300 and 159919) which includes the largest capitalization stocks in China stock market. In this research we focus on the ETF options traded at Shanghai Stock Exchange, underlying 50ETF and 300ETF.

Shanghai 50ETF (510050) was established on December 30, 2004 and the corresponding options start to trade on February 9, 2015. 50ETF option is the first exchange-traded ETF option in China market and is listed in Shanghai stock exchange. Huatai-Pinebridge Shanghai-Shenzhen 300ETF (510300) was established on April 5, 2012, and its option starts to trade on Dec 23, 2019. It was also the first cross-market ETF in China, which became one of the most effective and convenient choices for investors to invest in China economy.

Just like exchange traded stocks, ETF option trade for 4 hours per trading day. Each trading day 9:30-11:30, 13:00-15:00. 9:15 -- 9:25 and 14:57 -- 15:00 are the auction bidding time, when traders can undertake entrust to sign a bill.

50ETF (510050) turnover reached 1.14 billion shares at launch month and daily trading turnover is generally more than 2 billion nowadays. 300ETF (510300) sold 0.55 billion at launch month and daily trading turnover is usually around 1.5 billion.

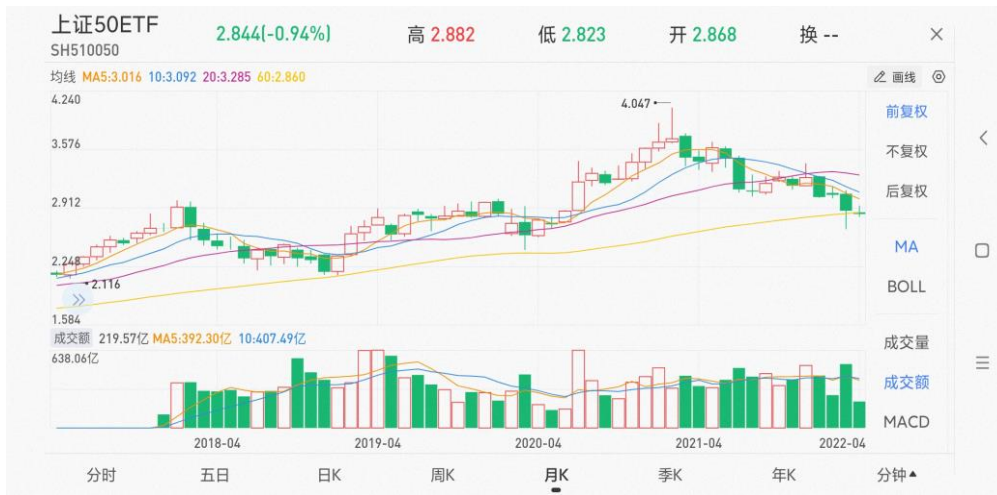

Figure1: 50ETF (510050) volume from 2018 Jan~ 2022 Apr, monthly; Source: https://xueqiu.com/S/SH510050

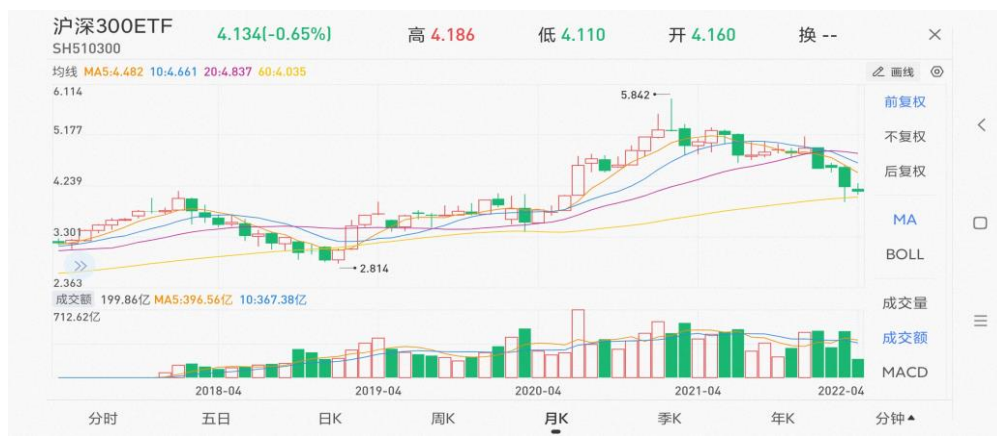

Figure2: 300ETF (510300) volume from 2012 Jan~ 2022 Apr, monthly; Source: https://xueqiu.com/S/SH510300

By the end of 2021, the outstanding value of ETF shares are 510300 56.886 billion RMB, 510050 69.135 billion RMB, and 159919 22.724 billion RMB.

In 2021, ETF option in Shanghai Stock exchange reached transaction volume of 1.097 billion contracts and 823.328 billion RMBs, corresponding to underlying value of 46.03 trillion RMBs, with corresponding daily average transaction volume of 3.388 billion RMBs.

All three ETF options are European-style options that can only be exercised on the exercise day, which is set to be the last Wednesday of every month (postpone to the next trading day if it happens to be a holiday).

## 2. Related theories for volatility-based trading:

Previous research has shown that some investors have high preferences on lottery-like investments with extreme positive returns, and the buy side of options has lottery characteristics of limited downside of whole loss of principles and theoretically unlimited return potentials. Another factor that drives up the price of out-of-money put option is that large institutions often need to hedge extreme downside systematic risk and thus put options on index are often purchased.

In that sense, option premiums would be driven up by those factors and become overvalued.

Under the current literature, European option value can be decomposed into intrinsic value, i.e., the value of option if exercised immediately, and time value, which is the traded option price minus the intrinsic value of the option.

Under classical option pricing models such as the Black-Scholes-Merton model, the time value of options comes from the possibility that the underlying would move in a favorable direction and increase the final payoff of the options. For example, an out-of-the money option's value all comes from the potentials of the option end up becoming in-the-money.

In particular, BSM model [2], [3] assumes that underlying move like random walks and options can be priced under the corresponding risk-neutral probability. The implications of BSM model suggest that options can be replicated by holding and dynamically adjusting certain numbers of underlying if the volatility keeps constant, which is the basis for many hedging activities for options.

By Black-Scholes Partial Differential Equation:

$$\frac{\partial V}{\partial t} + rS\frac{\partial V}{\partial S} + \frac{1}{2}\sigma^2 S^2 \frac{\partial^2 V}{\partial S^2} - rV = 0.$$

Formula1: Black Scholes Partial Differential Equations

On the above notation, V is portfolio value, r is risk-free interest rate (or cost of financing), S is the underlying value, and sigma the underlying volatility. In detail, the implication of this equation is that delta neutral portfolio (including options) will generate risk-free return, because the value impact of gamma (the second partial derivative of portfolio value on underlying price) induced by underlying price movement is removed by the change of time value of the portfolio (theta).

In reality, option prices are formed by continuous trading activities of buys and sells of different contracts sourced by supply and demand. Future realized volatility is always unknown and constant volatility is rare. A commonly used concept in understanding option price is implied volatility. For any price of the option, given other inputs of the BSM model which are more clearly defined, implied volatility can be back deducted for the current option price. When discrepancies appear between implied volatility and realized volatility, the implications of perfect hedging of gamma and theta will not hold. Also, option liquidity providers often need to hedge the position and the hedging cost is included in the form of increased option premium, also reflected as over-valued implied volatilities of options.

## 3. Trading Strategy

Based on previous discussions, a simple short-volatility strategy can be formed by dela-hedging such that when realized volatility is smaller than implied volatility, the strategy can make money by reaping the premiums reflecting the over-perceived volatilities from various sources. Besides, the Shanghai Stock Exchange has exempted fees for the sell side of the ETF options.

We take the daily historical data [11] of ETF options from initiation and construct the below short-volatility strategy. The data can be obtained from JoinQuant and Tushare financial data platform [11]. In detail, the portfolio is split into 2 equal parts at the beginning of each trading day and take short positions of calls and puts with the most time value decay (approximated by time value divided by duration) and closes all positions at the end of every trading day. To avoid extreme volatility and the increase of margin deposits, we don't consider contracts on the exercise day. The maximum position is set to be 90% of portfolio value and transaction cost is set to be 2 RMB/contract, where the fee from Shanghai Stock Exchange is 1.6 RMB/contract and a reasonable additional fee of 0.4 RMB/contract from the broker. Our preliminary results of this strategy are shown below.

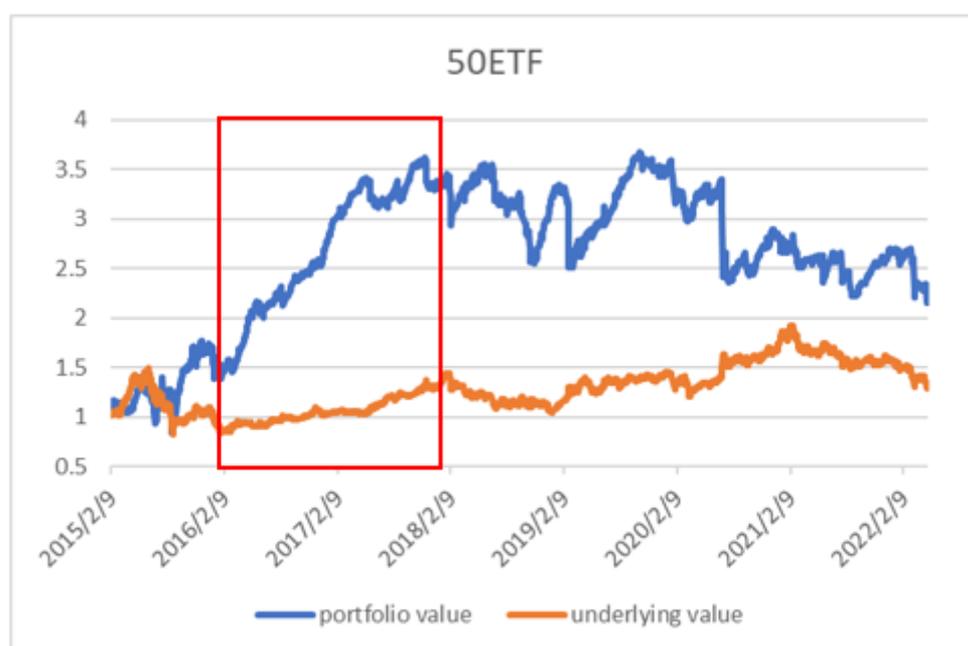

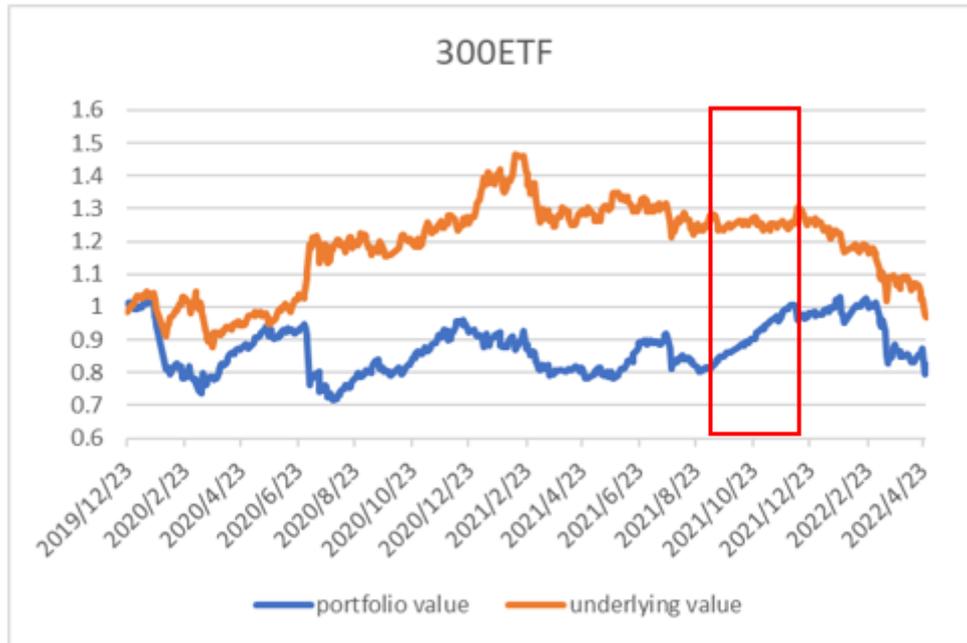

Figure3: short-volatility portfolio value and underlying price trend for 50ETF and 300ETF options

Aligned with previous theoretical discussion, the short-volatility strategy makes preferable returns in stable condition and drawbacks in volatile or trend market. Another point to note is that before 2018 the strategy delivers pretty good returns on 50ETF but after 2018 the performance start to deteriorate and also for the 300ETF option which starts to trade after 2019. The phenomenon could indicate a market condition shift where after 2018 the stock market in China become more volatile, and the recent development of quant PE funds in China could also contribute to a better volatility pricing such that volatilities are no longer over-priced and profits of single short-volatility strategy not that easy to capture anymore.

# 4. Volatility modelling and improved strategy

Now as we have discussed the strategy performs well in stable markets and bad in volatile markets, we could adjust the direction and exposure accordingly to forecasts of volatility and further improve from simple short-volatility strategy to volatility-based trading strategy. In this section we will discuss our volatility modelling in detail.

## Volatility Description [4]

Volatility refers to the degree to which asset prices fluctuate, similar to the concept of the standard deviation of the random variable in probability theory. Volatility cannot be observed directly, but some characteristics of volatility can be seen from the return on assets

To build a general model of volatility over time, we define volatility as the conditional standard deviation of return.

Let $r$ be the logarithmic rate of return based on a certain time unit of an asset at time $t$. $\{r_t\}$ series is considered to be coherent or low-order autocorrelation, but not independent time series.

Unitary volatility model is a kind of model that depicts return rate, which is self-irrelevant or low-order autocorrelation, but is not independent.

Let $F_{t-1}$ represent all historical information of the return rate up to time $t-1$, and consider the conditional mean and conditional variance of $r_t$ under the condition of $F_{t-1}$

$$\mu_t = E(r_t|F_{t-1}); \quad \sigma_t^2 = Var(r_t|F_{t-1})$$

Formula2: $r_t$ in $F_{t-1}$ conditional mean and conditional variance

$r_t$ can be decomposed to: $r_t = \mu_t + a_t$, while $\{a_t\}$ is the irrelevant white noise sequence, assuming $E(a_t|F_{t-1}) = 0$, it can be concluded from the above that:

$$\sigma_t^2 = Var(r_t|F_{t-1}) = Var(a_t|F_{t-1}) = E(a_t^2|F_{t-1})$$

Formula3: Volatility Formula

$\sigma_t$ means volatility, the conditional standard deviation of the return.

# Common Volatility Model [5]

## Moving Window and Volatility Momentum Method

The sliding window method assumes that the volatility level for the next N days is the same as that for the past N days. Therefore, as long as we estimate the historical volatility in the past, we can take the historical volatility as the predicted volatility in the future. Based on this concept we developed the volatility momentum strategy, as high volatilities are often followed by high volatilities based on previous financial time series research. Some new momentum strategies are defined as follows: we look back at the past 2n+1 days and check their daily return volatilities. If the majority of daily volatility has passed the volatility threshold, we go long volatility and vice versa we go short volatility. Theoretically, the threshold is defined as the volatility where a long straddle would break even. In our implementation to simplify the computation process, we take the proxy of passing the threshold as whether a long straddle strategy would incur a profit or loss. Some experiment results are reported below:

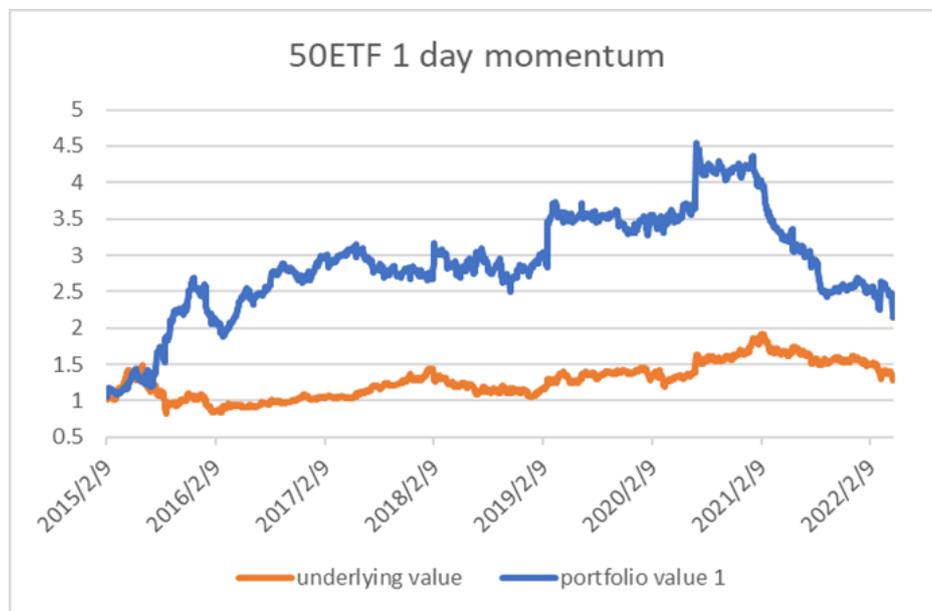

Figure4: 50ETF option 1 day volatility momentum strategy

| 50ETF strategy | underlying | original | momentum 1 | momentum 3 | momentum 5 | momentum 7 |
|---|---|---|---|---|---|---|
| final payoff | 1.2894 | 2.2281 | 2.1495 | 2.2692 | 0.9241 | 2.3762 |
| annualized return | 3.70% | 12.13% | 11.55% | 12.42% | -1.12% | 13.16% |
| Maximum value | 1.9172 | 3.6751 | 4.5407 | 3.3336 | 2.5537 | 4.2300 |
| Sharpe ratio (rf=2%) | 1.1519 | 2.2977 | 2.1676 | 2.3643 | -0.7081 | 2.5329 |

Chart1: 50ETF option volatility momentum strategies results.

From the figure, we can see again the strategy performs well for some time periods and not so satisfactory for the others, indicating that such once and for all predefined models are not likely to continuously make money under all market conditions. More flexible and complex models could be used to increase the prediction of realized volatilities.

**Exponentially Weighted Moving Average (EWMA)**

The exponential weighted moving average model taking the square of the yield of the latest period and the weighted average of the variance of the previous period to obtain the variance of the current period. This method is simple to use and easy to understand, but not sensitive enough. It is best to remove the outlier when forecasting future volatility, but the EWMA model simply avoids this problem by assuming that the impact of the event is exponentially decreasing.

In addition, this model does not take into account the market environment in which recent volatility estimators are located. For example, the EWMA model ignores the phenomenon that high volatility is often followed by low volatility, and the predicted value of any day is the same, which is not consistent with the actual situation.

**Generalized Auto-Regressive Conditional Heteroskedasticity (GARCH)**

The GARCH model introduced the long-term mean variance level term of the expected response, which solved the problem that EWMA could not achieve mean regression of volatility. In actual option trading, GARCH(1,1) model is often used to predict volatility:

$$\sigma_t^2 = \gamma V + \alpha r_{t-1}^2 + \beta \sigma_{t-1}^2$$

Formula4: Simplified GARCH(1, 1) model

$V$ stands for long-term mean of variance, and the model can also be extended to GARCH(p,q). Using GARCH model, we can obtain a term structure of future volatility, which gradually tends to the long-term mean.

GARCH model cannot explain the negative correlation between asset return and return fluctuation. GARCH(p, q) model assumes that the conditional variance is a function of the square of the lag residual, which means the sign of the residual does not affect the fluctuation. However, in actual trading and empirical research, it is found that volatility tends to increase when negative news appears, when expected asset returns will decline, volatility tends to decrease when there is good news. GARCH(p, q) model cannot explain this phenomenon.

The predication of future volatility could also be regard as a classification problem, we would like to try some machine learning method like decision tree or simple neural network to get the predication results. Meanwhile, in many cases we need not only a point estimate of future volatility prediction, but a probability distribution to predict the future, quantile characterization of observed volatility at different maturities of the same target may also needed.

Our team would choose GARCH model to present the modeling of historical volatility. And data source refers to [11]

## GARCH modelling

**Unit Root Test** [6]

Using ADF test to test unit root in time series.
In python, using "adfuller()" function from statsmodels.tsa.stattools to do the test.
Both 50ETF and 300ETF raw data has unit root, 1st differentiation is required to make the series stationary. After 1st differentiation, the series becoming stationary for follow-up work.

**Multicollinearity processing**

Using ACF and PACF [7] charts to figure tail cutting and trailing effect. It would help to determine the number of p and q in ARIMA(p, d, q) [8] model and GARCH(p, q) model.

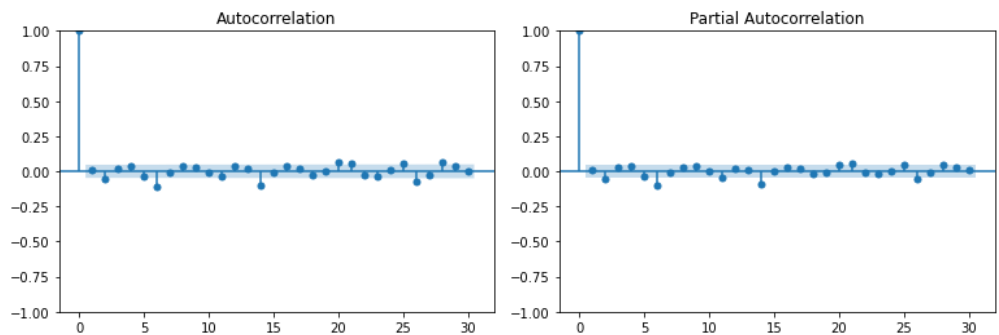

Figure5, 6: ACF and PACF chart of ETF50

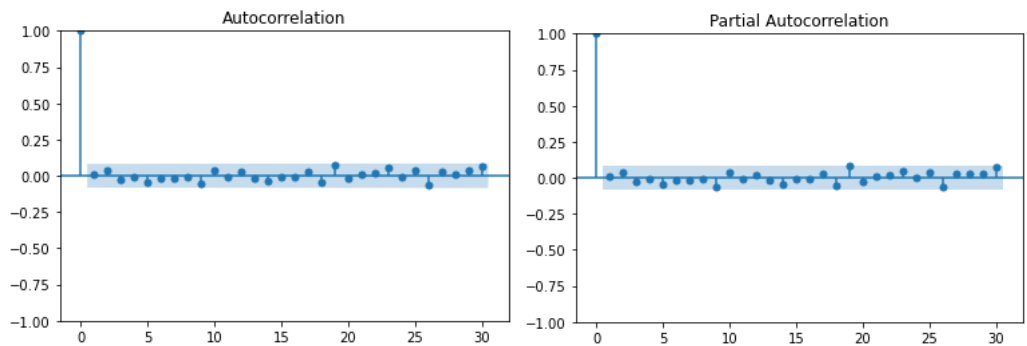

Figure7, 8: ACF and PACF chart of ETF300

In python, using the "tsa.arma_order_select_ic" function from "statsmodels.api" package could help quickly find out the best estimation of p and q for ARIMA model. The best p and q for ETF50 is $p = 6, q = 8$, for ETF300 the p and q equals to 0.

**GARCH Model Selection**

GARCH(p, q) model [1] could be presented as follow:

$$\sigma_t^2 = \alpha_0 + \sum_{i=1}^{q} \alpha_i y_{t-i}^2 + \sum_{j=1}^{p} \beta_j \sigma_{t-j}^2$$

Formula5: GARCH(p, q) model

GARCH(6, 8) for 50ETF model would provide 16 different parameters and make the model too complicated for further optimization and predication.

For most cases, GARCH(1, 1) would present good results with simplified model size and avoid of overfitting. The predication would be based on GARCH(1, 1) model for further steps.

```
Optimization terminated successfully.    (Exit mode 0)
            Current function value: -3389.2253419362714
            Iterations: 6
            Function evaluations: 19
            Gradient evaluations: 2
                    Constant Mean - GARCH Model Results
==============================================================================
Dep. Variable:            underly_close   R-squared:                       0.000
Mean Model:               Constant Mean   Adj. R-squared:                  0.000
Vol Model:                        GARCH   Log-Likelihood:                3389.23
Distribution:                    Normal   AIC:                          -6770.45
Method:              Maximum Likelihood   BIC:                          -6748.57
                                          No. Observations:                 1753
Date:                Wed, May 18 2022     Df Residuals:                     1752
Time:                        16:39:09     Df Model:                            1
                                  Mean Model
==============================================================================
                 coef    std err          t      P>|t|       95.0% Conf. Int.
------------------------------------------------------------------------------
mu         1.0118e-03  6.470e-04      1.564      0.118 [-2.564e-04,2.280e-03]
                              Volatility Model
==============================================================================
                 coef    std err          t      P>|t|       95.0% Conf. Int.
------------------------------------------------------------------------------
omega      3.3654e-05  1.296e-05      2.596  9.430e-03 [8.246e-06,5.906e-05]
alpha[1]       0.1000  2.331e-02      4.291  1.781e-05 [5.432e-02,    0.146]
beta[1]        0.8800  2.685e-02     32.777 1.244e-235 [    0.827,    0.933]
==============================================================================

Covariance estimator: robust
```

Chart3: GARCH output of ETF50

```
Optimization terminated successfully.    (Exit mode 0)
            Current function value: -790.0275977659214
            Iterations: 9
            Function evaluations: 70
            Gradient evaluations: 9
                    Constant Mean - GARCH Model Results
==============================================================================
Dep. Variable:            underly_close   R-squared:                       0.000
Mean Model:               Constant Mean   Adj. R-squared:                  0.000
Vol Model:                        GARCH   Log-Likelihood:                 790.028
Distribution:                    Normal   AIC:                          -1572.06
Method:              Maximum Likelihood   BIC:                          -1554.70
                                          No. Observations:                  566
Date:                Wed, May 18 2022     Df Residuals:                      565
Time:                        19:14:43     Df Model:                            1
                                  Mean Model
==============================================================================
                 coef    std err          t      P>|t|       95.0% Conf. Int.
------------------------------------------------------------------------------
mu         7.4172e-04  2.204e-03      0.337      0.736 [-3.577e-03,5.061e-03]
                              Volatility Model
==============================================================================
                 coef    std err          t      P>|t|       95.0% Conf. Int.
------------------------------------------------------------------------------
omega      3.5710e-04  2.463e-04      1.450      0.147 [-1.256e-04,8.399e-04]
alpha[1]       0.1687  8.929e-02      1.890  5.876e-02 [-6.247e-03,    0.344]
beta[1]        0.7556      0.118      6.392  1.636e-10 [    0.524,    0.987]
==============================================================================

Covariance estimator: robust
```

Chart4: GARCH output of ETF300

**Volatility Mapping**

By conducting the GARCH model, the parameter of predicting volatility is acquired,

the volatility could be calculated as:

$$\sigma_t^2 = \alpha_0 + \alpha_1 r_{t-1}^2 + \beta_1 \sigma_{t-1}^2$$

Formula6: GARCH(1, 1) for volatility prediction

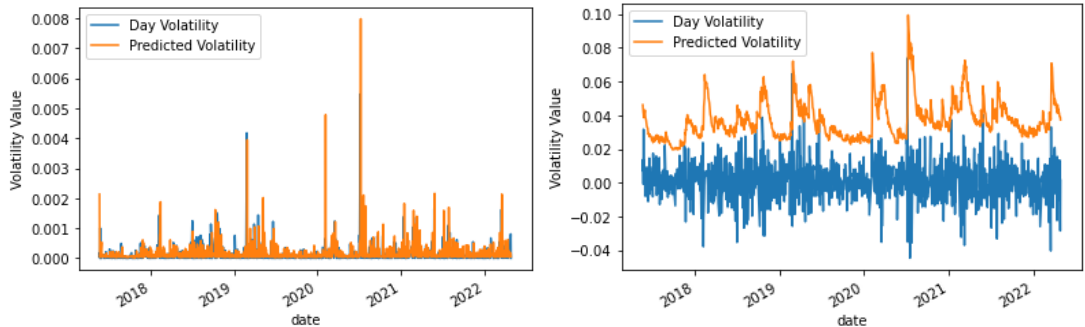

Figure9, 10: Volatility Prediction compare with real day volatility for 50ETF (Left for volatility square, right for standard volatility)

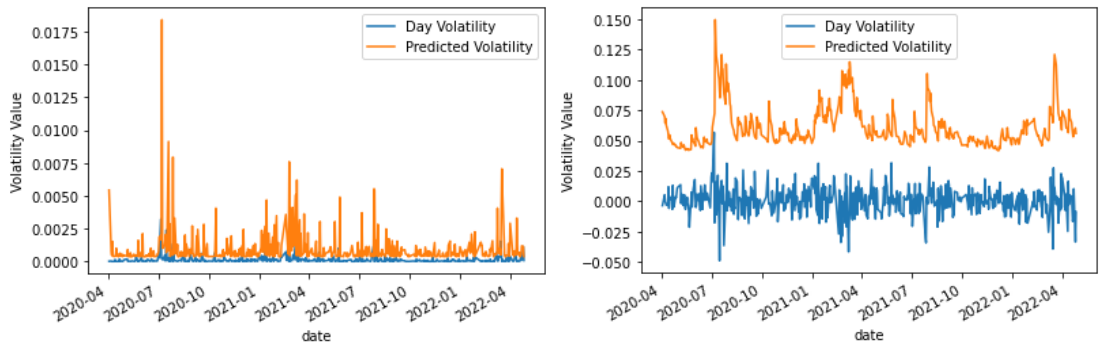

Figure11, 12: Volatility Prediction compare with real day volatility for 300ETF (Left for volatility square, right for standard volatility)

The volatility data needs be mapping to [-1, 1] section for adjusting the position in option market. Due to the unbalance distribution of volatility, using max-min [9] normalization method would result in many data very close to -1 and make the prediction results become a "mirrored image". The old strategy here is a simple short volatility strategy, when implied volatility is too high, traders should sell volatility and vice versa.

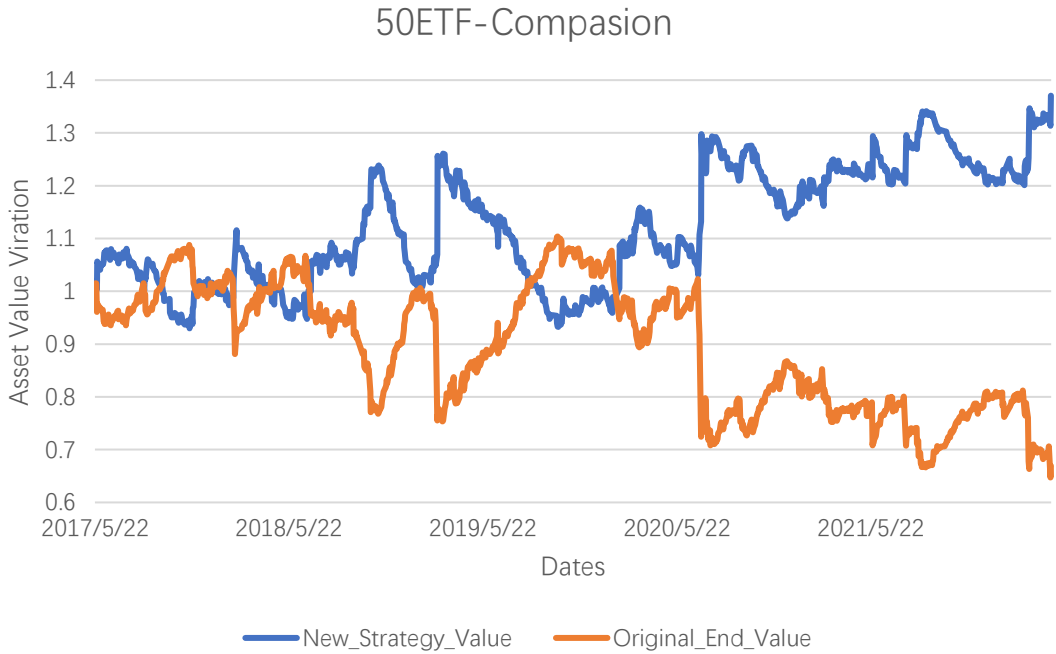

Figure13: Results after max-min normalization method for 50ETF

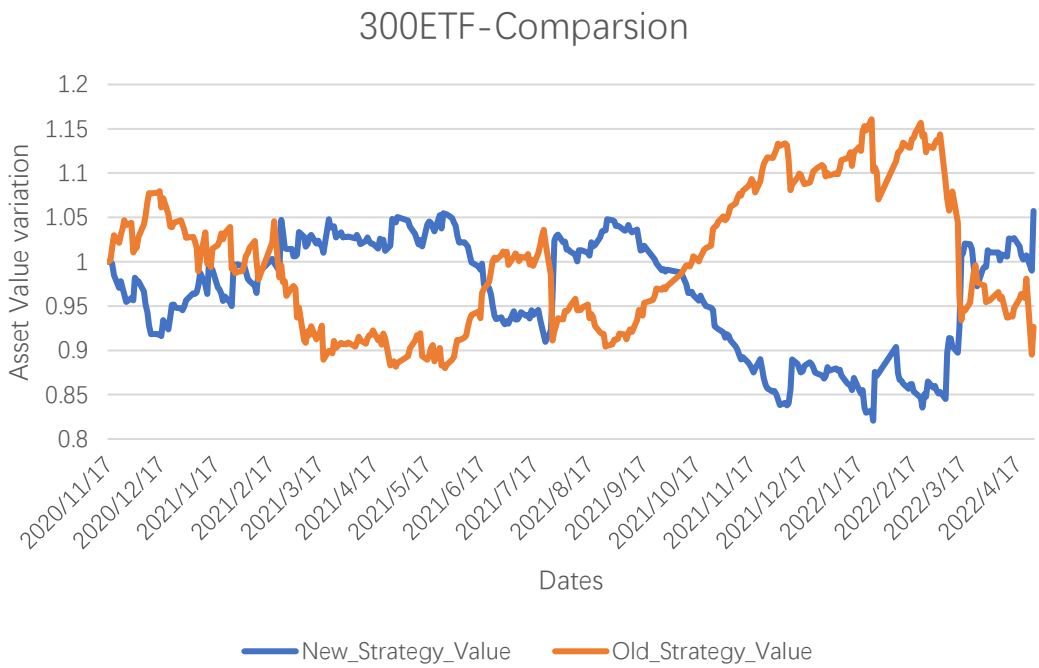

Figure14: Results after max-min normalization method for 300ETF

After many experiments, Z-scores [10] method would be more practical way for mapping, and this method works better for ETF50 data:

$$Mapped\ Value = \frac{x_{prediction} - \bar{x}}{\sigma}$$

Formula7: Z-score method for normalization

Another way for mapping is first using $x' = \frac{ln(x)}{ln(max(x_i))}$ , then use min-max normalization for mapping, this mapping method performs better in ETF300 data.

**Volatility Prediction**

$$New\ Strategy\ Value = Mapped\ Value * (End\_Value - Begin\_Value)$$

Using above formula to calculate the new strategy value and compare the new strategy with old strategy we would get following results:

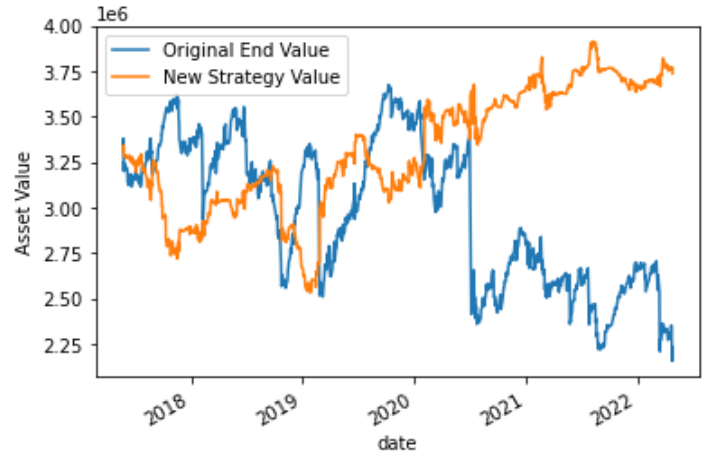

Figure15: 50ETF New strategy with old strategy (2017.5.22~2022.4.26)

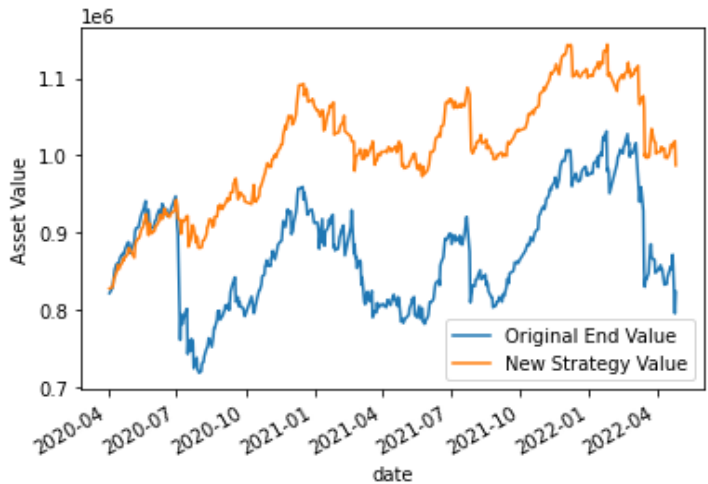

Figure16: 300ETF New strategy with old strategy (2020.4.03~2022.4.26)

If using volatility square in building the GARCH model, the higher the volatility, the trend could either be rapid raise or drop, the more volatile the data, the more potential nodes for profit.

The new strategy could be summary as the bigger the move the smaller the position, the small moves tend to hold positions.

**Volatility Optimization (Take ETF50 for example)**

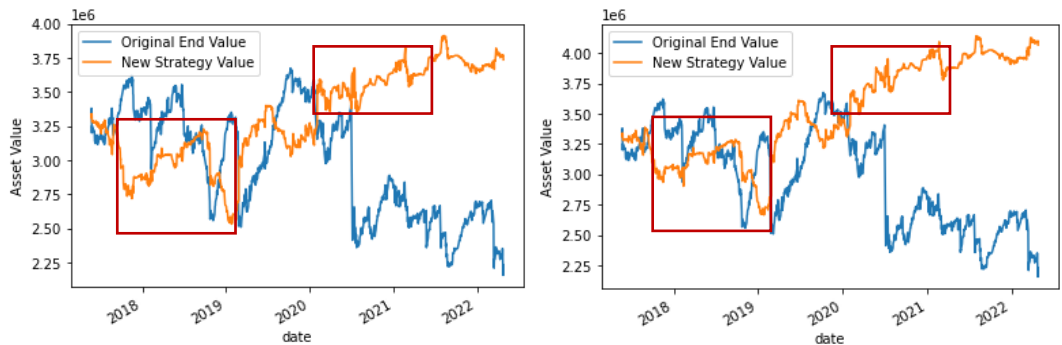

Figure17: Left one for original, Right one for optimization

The red box shows the right chart has less retracement when the ETF drop, it could be done by introducing a smoothing module, a simple way to perform the smoothing when drop is tracing the volatility few days ago, if the volatility value was negative for more than 3 days, volatility of today is more likely to be negative, using this strategy to optimize the position by reducing the position to a half or divided by square root of 2. And if the volatility value was positive for more than 3 days, be aware the loss of sudden drop, do not set the high position to speculate.

**Model Shortcoming (Take ETF50 for example)**

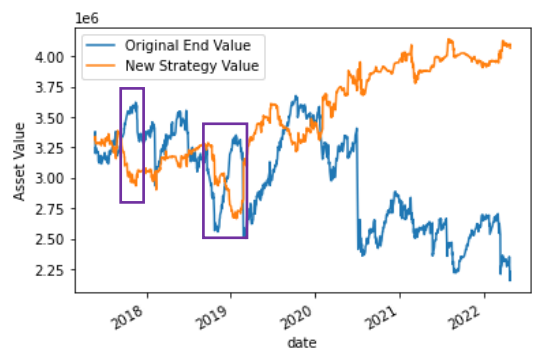

Figure18: Results after optimization

The purple box shows the new strategy goes the opposite direction compare with simple short volatility strategy. The processing of small fluctuations is not sensitive enough, an improved way is to collect more historical data to adjust the predicted volatility value, the model sensed the volatility but mislead the position because of lack of information of current time and doesn't adjust immediately.

# 5. Conclusion

To conclude, from the daily data of Chinese equity index ETF options, we found that a simple short-volatility strategy would deliver outstanding performance before 2018 but deteriorate afterwards. Under the Efficient Markets Hypothesis and traditional option pricing model such as the BSM model, the return on any investable portfolios would not provide risk-adjusted returns. This could indicate the options were over-priced in China in the earlier years and with the recent rapid development of quant funds, such mis-pricing become less obvious. Still, with volatility forecasting, paper trades of long and short straddles or strangles of options could still provide attractive returns compared to the benchmark of buy and hold of the underlying. In our research, volatility forecasts models such as GARCH show potentials to improve the volatility-based trading strategy, with the idea of adjusting positions according to the predicted volatilities. The updated model could beat the simple short volatility model for recent years with market condition in more uncertainty, while it remains the space for improvement in extreme volatile and depressed markets.

# Reference(Format request for updating)

| Variable | Description |
| --- | --- |
| Date | Trade date |
| Begin_value | The beginning value of asset of trade date |
| End_value | The end value of asset of trade date |
| Call_contract | Call contract type |
| Put_contract | Put contract type |
| Num_call | The number of call option exercised |
| Num_put | The number of put option exercised |
| underly_open | The beginning value of underlying assets |
| underly_close | The end value of underlying assets |